\documentclass[useAMS]{mn2e}
\usepackage{lscape}
\usepackage{amssymb}
\usepackage{rotating}

\def\rcore{\hbox{R$_{\rm core}$}}
\def\tr{\hbox{t$_{\rm relax}$}}

\def\trelaxnow{\hbox{t$_{\rm rel, current}$}}
\def\trelaxreal{\hbox{t$_{\rm rel, real}$}}
\def\msun{\hbox{M$_\odot$}}

\def\cm3{\hbox{cm$^{-3}$}}

\input psfig.sty
\input epsf.sty
\usepackage{graphicx}
\usepackage{epsfig}
\title[Expansion of cluster cores]
{The Early Expansion of Cluster Cores}
\author[Bastian et al.] {N. Bastian$^{1,2}$, M. Gieles$^3$,
  S.P. Goodwin$^4$, G. Trancho$^5$, L. J. Smith$^{2,6}$, \newauthor I.
  Konstantopoulos$^2$, Yu. Efremov$^7$\\ 
$^1$ Institute of Astronomy, University of Cambridge, Madingley Road, Cambridge, CB3 0HA, UK\\
$^2$ Department of Physics and Astronomy, University College London,
  Gower Street, London, WC1E 6BT, UK\\ $^3$ European Southern Observatory, Casilla 19001, Santiago 19, Chile\\
$^4$ Department of Physics and Astronomy, The University of Sheffield,
  Hicks Building, Hounsfield Road, Sheffield S3 7RH, UK\\ 
$^5$ Gemini Observatory, 670 N. A'ohoku Place, Hilo, HI 96720, USA\\
$^6$ Space Telescope Science Institute and European Space Agency, 3700 San Martin Drive,
  Baltimore, Maryland 21218, USA\\
$^7$ Sternberg Astronomical Institute of Moscow State University, Universitetsky
 Prospect, 13, Moscow, 119899, Russia\\
}
\date{Accepted. Received; in original form}
\pagerange{\pageref{firstpage}--\pageref{lastpage}}
\pubyear{2005}
\begin{document}
\maketitle
\label{firstpage}
\begin{abstract}
The observed properties of young star clusters, such as the core
radius and luminosity profile, change rapidly during
the early evolution of the clusters.  Here we present observations of 6
young clusters in M51 where we derive their sizes using HST imaging
and ages using deep Gemini-North spectroscopy.    We find evidence for
a rapid expansion of the cluster cores during the first $20$~Myr of
their evolution.  We confirm this trend by including data from the
literature of both Galactic and extra-galactic embedded and young
clusters, and possible mechanisms (rapid gas removal, stellar evolutionary mass-loss, and internal dynamical heating) are discussed.   We explore the implications of
this result, focussing on the fact that clusters were more concentrated
in the past, implying that their stellar densities were much higher
and relaxation times (\tr) correspondingly shorter.   Thus, when
estimating if a particular cluster is dynamically relaxed, (i.e. when
determining if a cluster's mass segregation is due to primordial or
dynamical processes), the current relaxation time is only an
upper-limit, with \tr~likely being significantly shorter in the past.

\end{abstract}

\section{Introduction}

The early evolution of stellar clusters and aggregates has a rich
variety of physical processes at work including: stellar formation and
evolution, gas inflow and outflow, stellar
feedback and turbulence, the merging of stellar clumps and possibly,
stellar interactions.   The combination and effective efficiencies of
these processes determine if the cluster, or part thereof, 
becomes/remains bound or if it forms an unbound loose aggregate of stars
which will slowly blend into the background field population.  These 
processes leave their mark on the cluster, in the 
size (core or effective radius), mass, profile
shape, and possibly on the stellar mass function.

This work is a continuation of our previous investigations on the
implications of rapid residual gas expulsion (RGE) on the
survivability and properties of young clusters (Bastian \&
Goodwin~2006; Goodwin \& Bastian~2006).  In previous papers we have
explored the evolution of the luminosity profile of the clusters as
well as their dynamical state.  Both were found to be highly variable
which led us to conclude that the observed properties of young
clusters were merely snapshots in their evolution and should not be
regarded as their final properties.  One general prediction from our
models, as well as other models of RGE (e.g. Goodwin~1997, Kroupa \&
Boily~2002), is that the cluster will expand in response to the loss
of the residual gas, the exact amount of which will depend on the
(effective) star-formation efficiency\footnote{Goodwin~(2008) reiterates
  that it is not the star formation efficiency {\it per se} that is the
  critical factor in determining the effect of RGE, rather the
  dynamical state of the cluster at the onset of RGE which can be
  parameterised as an effective star formation efficiency.}.

In the current work, we investigate the evolution of core radii for a
sample of young clusters.  The sample is partly composed of a small
survey of young (age $< 30$~Myr) clusters in M51 for which we use high
S/N ($>100$) optical spectra in order to derive their ages, and
HST-ACS imaging to derive their core radii.  We supplement this sample
with clusters taken from the literature, composed of both embedded and
open clusters in the Galaxy, as well as massive extra-galactic
clusters.  These datasets are designed to complement the study of Mackey
\& Gilmore~(2003) who derived the core radius for 63 clusters in the
LMC/SMC and found a strong relation between the core radius of a
cluster and its age (as first found by Elson~1991), in the sense 
that older clusters have a wider spread of core radii than young clusters.

The core radius of a cluster is a particularly interesting parameter
as it is largely responsible for setting the timescale over which the
cluster evolves dynamically.  For a given mass, it is the core radius
which will set the core relaxation timescale and determine how quickly
dynamical mass segregation proceeds and whether or not stellar mergers
are likely to take place (Freitag et al.~2006), assuming that the underlying stellar IMF is sufficiently broad (G{\"u}rkan, Freitag, \& Rasio~2004).   The core radius of
the cluster is expected to increase during the first few 10s of Myr due
to three main effects. Firstly, from stellar evolution in which the most massive
stars lose mass (this effect is heavily amplified if the core is
mass-segregated\footnote{i.e. the most
massive stars are found preferentially in the centre of the cluster,
more than would be expected from randomly sampling from the stellar IMF in a centrally concentrated
profile.} - e.g. Mackey et al.~2007). Secondly, due to the
expulsion of gas left over
from the non-100\% efficiency star-formation process (RGE, see Goodwin \&
Bastian~2006; Goodwin~2008 and references therein).  Thirdly, dynamical heating of the core through 'dark objects' (i.e. black holes and neutron stars) interactions with lower-mass stars (e.g. Merritt et al.~2004, Mackey et al.~2007,2008).  All three effects 
are understood relatively well
theoretically (see the recent review by Baumgardt \& Kroupa~2007) and
all are likely to play a large role.  The goal of the present paper
is to test this theoretical framework with observations.   In addition to the above effects, external perturbations such as interactions with GMCs and other clusters, disk shocking and spiral arm passages are expected to also heat the cluster, causing them to expand (e.g.~Gieles et al.~2006,2007)

This paper is organised in the following way.  In \S~\ref{sec:data} we
present the observations and numerical techniques.  In
\S~\ref{sec:parameters} we describe in detail the methods employed to
derive the age and core radius of each of the clusters in the M51
sample.  In \S~\ref{sec:radius-age} the age-core radius relation is
discussed using the M51, Galactic and other extra-galactic cluster
samples and in \S~\ref{sec:causes} possible mechanisms are summarised.
We discuss the results and implications in \S~\ref{sec:discussion} and
summarise the results in \S~\ref{sec:conclusions}.

\section{Observations}\label{sec:data}

The spectroscopic observations were taken the nights of May 25-26th,
2006, using the GMOS spectrograph on the Gemini-North telescope in long slit mode (PI
Bastian, GN-2006A-C-9).  We used a slit width and length of 1.0" and 5.5' respectively, and the B600
grating to achieve a resolution of $\sim150$~km/s.  We chose three slit positions which were based on the catalogue of young cluster complexes in M51 by Bastian et al.~(2005), and we use their naming
convention throughout this paper (the cluster positions in the galaxy can be found using Fig.~1 in Bastian et al.~2005).  For each slit
position, we obtained two 1800s exposures, which were centred on 508 and
512~nm.  For all observations the seeing was in the 70th percentile (i.e. better than 0.8").  The data were flat-fielded, bias subtracted, wavelength
calibrated, extracted, and combined using standard Gemini/IRAF
software.   

\begin{figure}
\includegraphics[width=8.5cm]{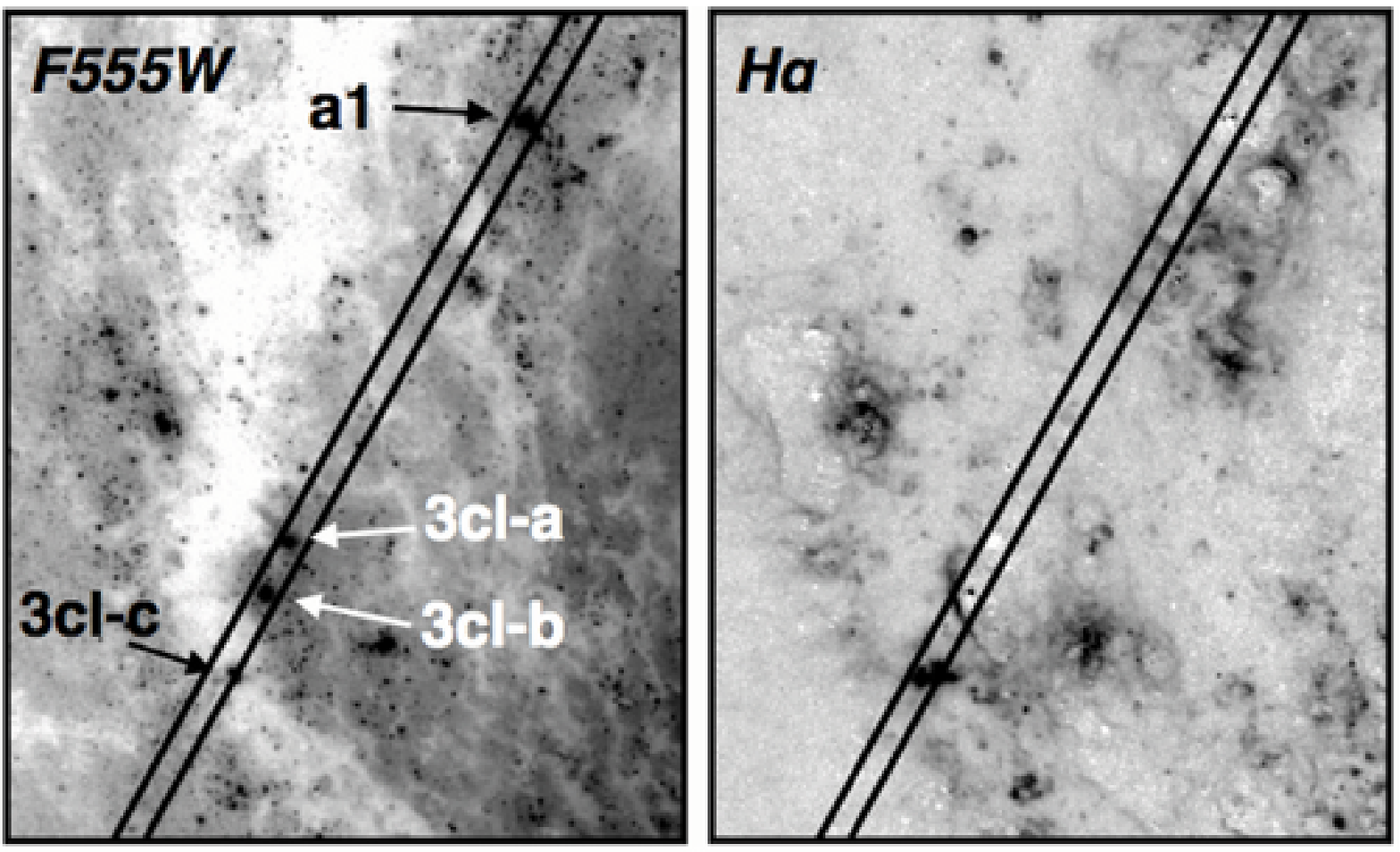}
\includegraphics[width=8.5cm]{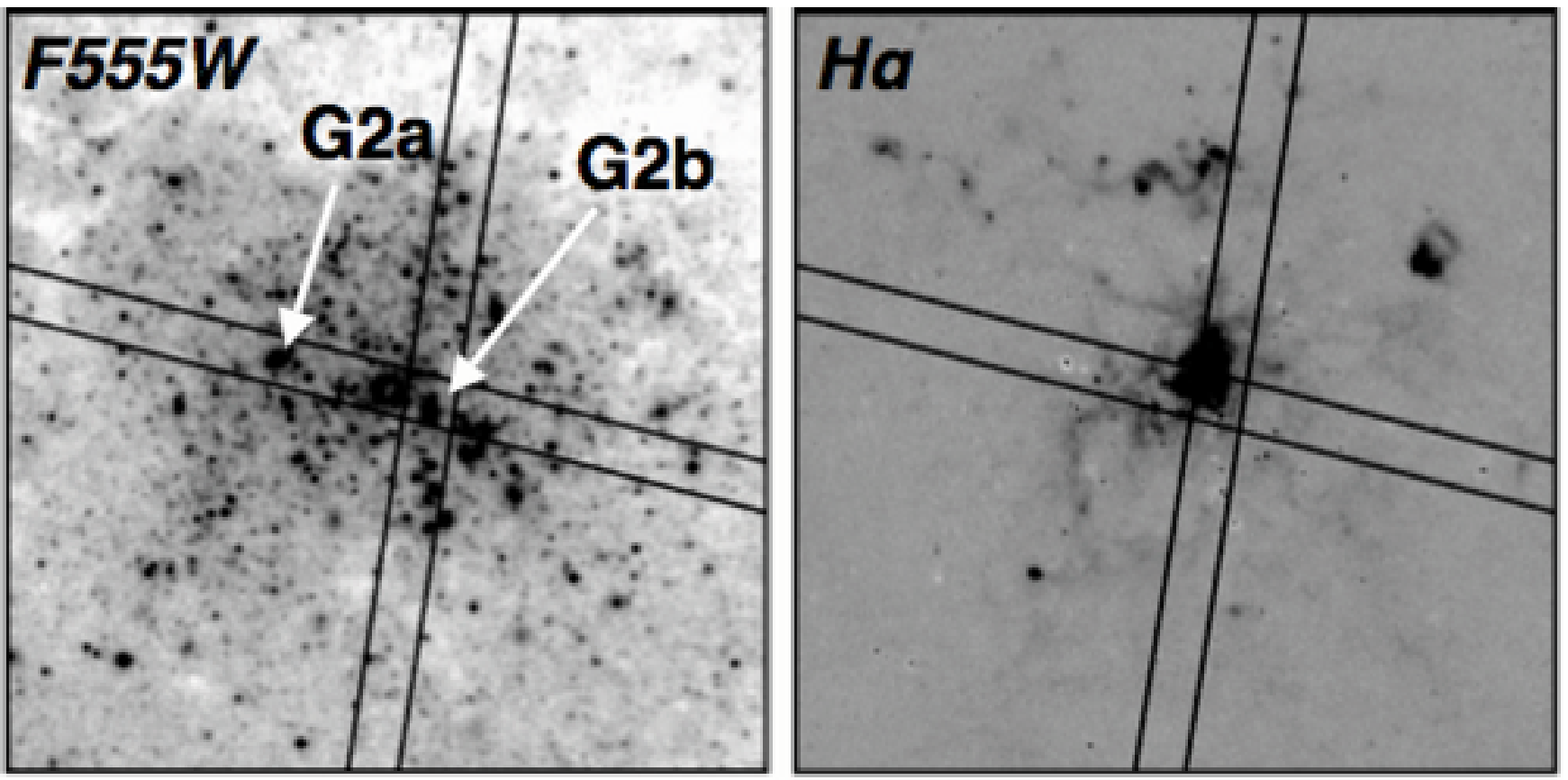}
\caption{{\bf Top:} Slit positions superimposed on {\it HST F555W} and
  $H\alpha$ (continuum subtracted) images of four of the clusters in
  the sample.   Each image is 27.5'' by 33.75'', corresponding to
  1.12~kpc by 1.375~kpc.  North is up and east is to the left.  {\bf
  Bottom:} Slit positions superimposed on {\it HST F555W} and
  $H\alpha$ (continuum subtracted) 
  images of the two clusters within the complex G2.  Each image is
  $\sim610$~pc on a side.  All images are shown in negative scaling, where dark shading refers to greater intensity and light regions are places of low intensity or high extinction.} 
\label{fig:a1-slits}
\end{figure}

\begin{figure}
\includegraphics[width=8.5cm]{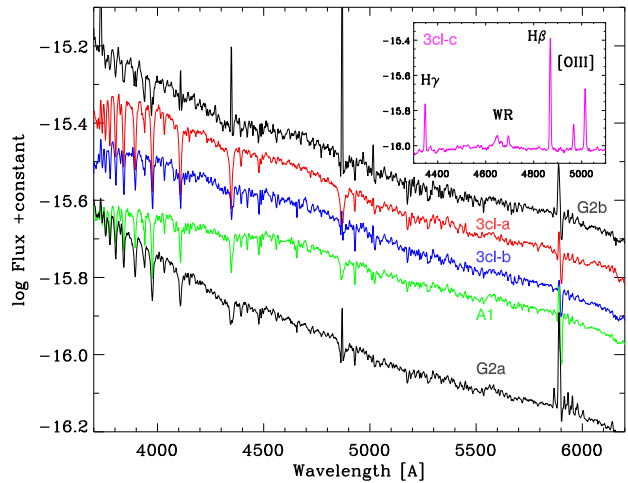}
\caption{Spectra of the six clusters observed in M51.  Differences in
  the continua of the spectra are due to different amounts of
  extinction and also due to the time and angle of the different
  slits, as no atmospheric dispersion correction was applied.  Strong
  emission lines in cluster 3cl-c are labelled in the inset.} 
\label{fig:spectra}
\end{figure}

Since the slit positions were
chosen to cover multiple complexes in the same pointing, the positions
were independent of paralactic angle.  As such, we have not corrected
for wavelength dependent slit losses, which accounts for some of the
observed differences in the spectral shapes of the clusters.
The slit and cluster positions are shown in Fig.~\ref{fig:a1-slits},  their coordinates are given in Table~\ref{table:fits} and the spectra are shown in Fig.~\ref{fig:spectra}.

Each slit contained one to four clusters with individual clusters a1 and G2b observed during two different pointings (i.e. for a total of four exposures for these clusters).  The spectra show features common to young stellar populations, namely a combination of emission lines and strong Balmer absorption lines.

The structural parameters of the clusters were derived using
HST-ACS-WFC observations (F435W, F555W, and F814W).  These
observations were taken as part of the Hubble Heritage Project in
January 2005 (proposal ID 10452, PI: S. V. W. Beckwith) and the data
reduction and processing are described in detail in Mutchler et
al.~(2005).  Throughout this paper we will use the standard B, V, I
notation to discuss the colours of the clusters, however we note that
no transformation has been applied. 

We adopt a distance to M51 of 8.4~Mpc (Feldmeier et al.~1997). 

\section{Measuring Parameters}
\label{sec:parameters}

\subsection{Ages}

\begin{figure}
\includegraphics[width=8.5cm]{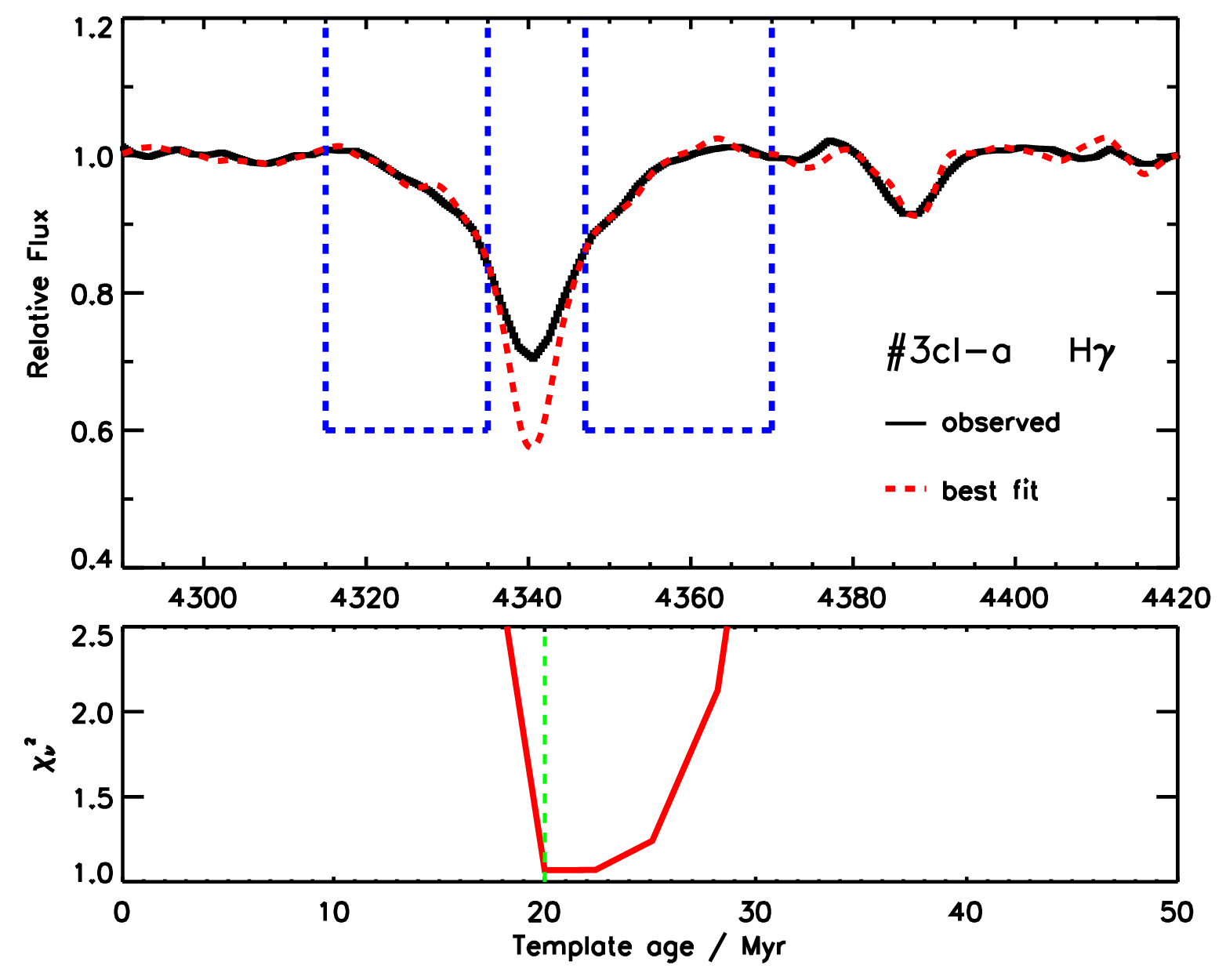}
\includegraphics[width=8.5cm]{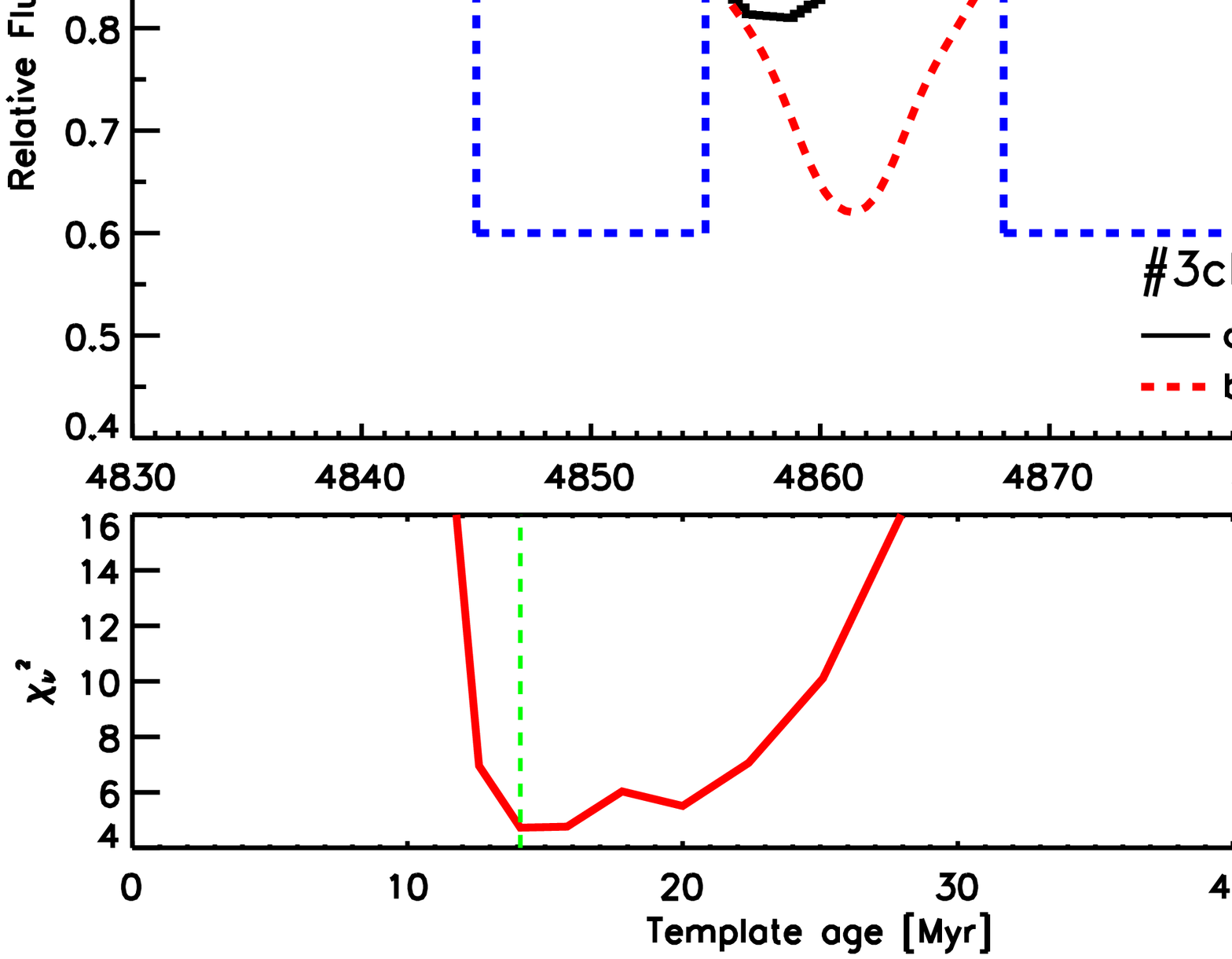}
\caption{The observed spectra and best fitting template spectra for
  cluster 3cl-a around the H$\gamma$ (top) and H$\beta$ (bottom)
  absorption lines.  The (blue) dashed boxes represent the spectral
  wavelength region used in the fits, where the centre of each line
  was not used due to a clear emission component.  The lower plot in
  each panel shows the $\chi_{\nu}^2$ result for each model fit,
  along with the best fitting model (marked as a vertical dashed line).} 
\label{fig:example}
\end{figure}

Optical spectra are a powerful way to derive accurate ages for young clusters (e.g. Trancho et al. 2007a,b).  For the present study we adopt the
technique presented in Konstantopoulos et al.~(2008), and we refer the
reader there for the details of the method.  In short, the method compares
the detailed line profile shapes of the H$\gamma$ and H$\beta$ lines
with the Gonzalez-Delgado et al.~(2005) simple stellar population
models which have been degraded in resolution to match the
observations (we have used a Salpeter stellar IMF, and solar
metallicity tracks).  The comparison between the model and observed
spectra is done on rectified spectra in two bands which straddle the
line.  The centre of the line is avoided in order to minimise
contamination from any underlying emission component.  This comparison
is done for model ages between 4~Myr and 10 Gyr and the model with the
lowest reduced $\chi^2$, $\chi_{\nu}^2$, is selected.  The range of acceptable model ages was determined by comparison of the models and observations by eye.  In particular, we compared the line width and overall profile fit, including small features in the profile which were seen in the observed spectrum as well as the best fit model.  An example of the procedure (3cl-a), is shown in Figure~\ref{fig:example}. 

We have also fitted clusters a1, g2a, 3cl-a, and 3cl-b with SSP model
tracks with Z=0.008, for which find good agreement with the solar
metallicity fits.  The results are given in Table~\ref{table:fits},
however due to the good agreement, we will adopt the ages derived
assuming solar metallicity throughout the paper.  We note that cluster a1 is found in the center of an H$\alpha$ bubble which is  approximately 80~pc in radius.  This may argue for a higher age, namely that found using the Z=0.008 models, but for consistency we adopt the younger Z=0.02 results for cluster a1.

For cluster 3cl-c, the lack of any absorption lines in the observed
spectrum make this technique unfeasible.  However, this cluster
appears to be deeply embedded in a dust lane and has strong emission
associated with it (see top panel of Fig.~\ref{fig:a1-slits}), which
points to a very young age ($<<10$~Myr).  Additionally, the 'blue
bump' is clearly observable in the spectrum at $\sim4650$~\AA~ which is
a feature normally attributed to the presence of Wolf-Rayet stars.  Such stars
have very short lifetimes and their presence in the cluster indicates
an age between 2 and 5 Myr (see Crowther 2007 and references therein). 

Cluster G2b appears similar to 3cl-c in the lack of strong absorption
lines.  It does not, however, show any strong Wolf-Rayet features in
the spectrum.  Due to the proximity of this cluster to the
H{\sc ii} region seen in the right panel of Figure~\ref{fig:a1-slits},
we associate this cluster with a young age, namely $5\pm2$~Myr.

\subsection{Structural Parameters}

In order to determine the structural parameters of the clusters we
used the {\it ISHAPE} algorithm (Larsen~1999).   We empirically
derived the PSF from bright isolated stars in the field of view. 

\begin{itemize}
\item {\bf 3cl-a, 3cl-b, a1:} These three clusters are extremely
  bright in all three bands (BVI) and hence we were able to have the
  index of the Elson, Fall \& Freeman (1987, hereafter EFF) profile 
  as a free parameter.  A fitting radius of
  15 pixels ($\sim30$~pc) was used.  The errors were estimated from the
  standard deviation between the B, V, and I-band fits.   We have also estimated the errors in the fits using version 0.93.9beta of {\it ISHAPE}, which calculates the errors, including correlations between the parameters, and find errors slightly smaller than the standard deviation between the filters.

\item {\bf G2a:} No best fitting profile could be found, so we assumed
  an EFF profile and varied the index between 1 and 2.5 ($2 \le \gamma
  \le 5$), which are typical values for clusters in the LMC (e.g. Mackey \& Gilmore~2003).  We carried out the fits on all three bands (BVI) and took
  the average.  The error was estimated in the same way as the above clusters.

\item {\bf 3cl-c,G2b:} We used a fitting radius of 10 pixels due to
  contamination from nearby objects.  No clear best fitting profile
  could be found.  We put an upper limit on the size by fitting EFF
  profiles with indices between 1 and 2.5 ($2 \le \gamma \le 5$) and
  found cluster radii between 0 (unresolved) and 0.42~pc. 

\end{itemize}

One potential caveat in this method is that it implicitly assumes that the distribution of light within the cluster represents the underlying distribution of mass.   If these clusters are, however, severely mass segregated then the profile derived from the light will underestimate the actual core radius (since the light will be dominated by the most massive stars which are more concentrated than the lower mass stars - e.g. Gaburov \& Gieles~2008).  

\begin{table*}
\begin{centering}
{\scriptsize
\parbox[b]{7.5cm}{
\centering
\caption[]{The properties of observed clusters in M51}
\begin{tabular}{c c c c c c}
\hline
\noalign{\smallskip}
Cluster ID  & age$^a$  &$\chi^2_{\nu,\rm best}$$^{b}$ & age$^c$ & R$_{\rm core}$ & Coordinates \\ 
                    &  (Myr) &  & (Myr) & (pc)    &    (J2000)\\ 
\hline
a1 & 5 $^{8.9}_{4}$ & 4.1& 7.3 $^{10}_{4}$ &$0.63\pm0.10$ & 13:29:54.64   47:12:08.1 \\\
\\
3cl-a & 16.5 $^{25.1}_{12.6}$  & 1.5 & 20 $^{28}_{14}$ & $1.65\pm0.05$ & 13:29:55.59  47:11:50.9\\
\\
3cl-b & 5 $^{6}_{4}$ & 4.7 & 6 $^{4}_{10}$ & $1.02\pm0.33$ & 13:29:55.67  47:11:48.8\\
\\
3cl-c & 3 $^{5}_{2}$$^d$ & -- &  -- &0.38$^e$ & 13:29:55.81  47:11:45.6\\
\\
G2a& 6 $^{14}_{4}$ & 1.9& 10 $^{12}_{5}$ & $1.08\pm0.35$ & 13:29:43.31  47:11:38.8\\
\\
G2b & 5 $^{7}_{3}$  & -- &  -- &$ 0.42^e$ & 13:29:43.02  47:11:37.8\\
\noalign{\smallskip}
\noalign{\smallskip}
\hline
\end{tabular}
\begin{list}{}{}
\item[$^{\mathrm{a}}$] The best fit age is given (solar metallicity), along with the lower and upper limits as defined in the text.
\item[$^{\mathrm{b}}$]  $\chi^2$ of the best fitting template age for H$\beta$.
\item[$^{\mathrm{c}}$]  Same as for (a), but for Z=0.008.
\item[$^{\mathrm{d}}$]  Age based on the presence of Wolf-Rayet emission features in the spectrum.
\item[$^{\mathrm{e}}$]  Only an upper limit, as discussed in the text.

\end{list}
\label{table:fits}
}
}
\end{centering}
\end{table*}

\section{The core radius/age relation}
\label{sec:radius-age}

\begin{figure}
\includegraphics[width=8.5cm]{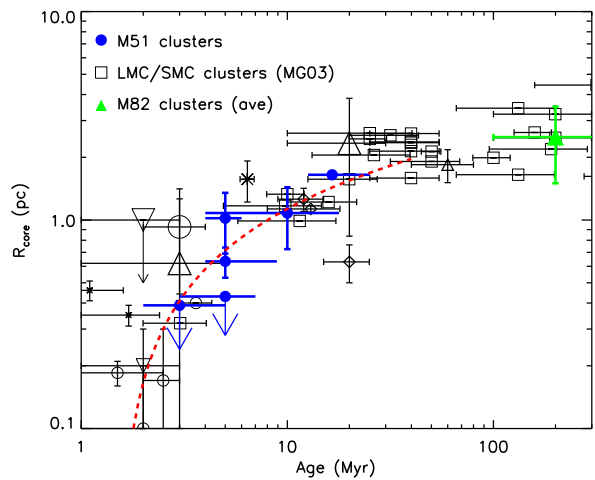}
\caption{The relation between core radius and age of M51 clusters
  (filled blue circles) and other clusters taken from the literature.
  The large symbols represent median values of cluster surveys, see
  text for details.  The dashed (red) line is a logarithmic fit to the
  data, done by eye ($R_{\rm core} [{\rm pc}]= 0.6 \times
  \ln{\rm(age [Myr])} - 0.25$).  
}
\label{fig:core-evo}
\end{figure}

Figure~\ref{fig:core-evo} shows the relation (filled blue circles)
between the derived core radius and age for the six clusters in M51.
There is a clear relation, with older clusters being larger than
younger ones. 

Young clusters are generally not found in isolation, but rather as
parts of larger complexes due to the hierarchy of star-formation
(e.g. Zhang, Fall, \& Whitmore~2001; Bastian et al.~2005).  As such,
we expect, and observe, many sources around the young clusters (e.g. in the
complex G2).  These additional sources may cause blending with the
clusters of interest, making them appear larger than they actually
are.  This bias, however, works in the opposite way to the observed trend
(that the younger clusters are smaller), hence the actual trend
may be stronger than we have observed. 

In order to check if the observed relation between age and core radius
is simply a reflection of an underlying mass-radius relation, we have estimated
the mass of each of the clusters.  For this we have compared the
observed BVI colours of each cluster to the {\it GALEV} simple stellar
population models, assuming solar metallicity and a Salpeter IMF
(Anders \& Fritze-v.~Alvensleben~2003).  We use the best fitting
spectroscopic age of each cluster and determine the cluster reddening
based on the deviation between the observed colours and those expected
at that age.  We then use the age dependent M/L ratios to estimate the
mass using the extinction corrected V-band flux. 

We find that clusters a1, 3cl-a, 3cl-b, and 3cl-c have similar masses
within a factor of two ($\sim0.7 - 1.3 \times 10^5$~\msun).  G2a and
G2b have similar masses (a few $\times 10^4$\msun), although G2a is at
least twice as large (core radius) as G2b.  Hence we conclude that
there is not any strong mass-radius relation present within this small
dataset. 

In order to understand the R$_{\rm core}$-age relation we searched 
the literature and found young clusters which have had their ages and
core radii measured.  We take only clusters which have had their ages
derived by either CMD fitting or spectroscopic age dating in order to
have as clean a sample as possible.  Mackey \& Gilmore~(2003) 
presented a large database of LMC/SMC clusters with accurate core
radii and ages,  these are shown as open squares in
Fig.~\ref{fig:core-evo}.  In the Galaxy there have been a number of
massive young clusters discussed, including NGC 3603,  Westerlund 1,
Westerlund 2, the Arches, and the Orion Nebula Cluster (compilation
taken from Brandner et al.~2007; however using an age of 1.5~Myr for
the ONC - Jeffries~2007),  NGC~2316 (Teixeira et al.~2004),
Trumpler~14 and DBSB48 (Ortolani et al.~2008).   
Some massive extra-galactic clusters have also been included, namely:
NGC~1569B (Larsen et al.~2008), NGC~5236-805 (Larsen \& Richler~2004),
NGC~6946-1447 (Larsen et al.~2001),  M82F (Bastian et al.~2007, and
references therein), and M82-A1 (Smith et al.~2006). 

In addition, we also include surveys of cluster systems.  The surveys
are included in Fig.~\ref{fig:core-evo} as large open symbols, where
the error bars on R$_{\rm core}$ represent the standard deviation of
all members and the symbols represent the median.  The Rosette nebula
(Roman-Zuniga et al.~2007) was included, which is a group of nine
clusters still in the embedded phase (age$\sim3-5$~Myr).  We include
the survey of embedded clusters by Lada \& Lada~(2003) (assigning an
average age of $3\pm2$~Myr).  From the Kharchenko et al.~(2005) catalogue
of open clusters we take the mean core radius of all clusters with
ages between 10 and 30~Myr (three clusters with estimated core radii
larger than 20~pc were excluded).   We have taken the mean values of
Johnson et al.~(2003) for young embedded radio detected clusters in
IC~4662 whose core radii were estimated to be less than 1~pc, with
adopted ages of 1-3~Myr.  Finally, we include all clusters in M82 with
ages between 100 and 200~Myr, from the recent study by Konstantopoulos
et al.~(2008b - in prep.). 

Figure~\ref{fig:core-evo} clearly shows that the all of the clusters follow
the trend observed in the M51 clusters -- core radii increasing with 
age.  The possible
causes of this, and the implications are discussed in the next
section. 

Such a relation between cluster size and age has been seen before,
albeit with smaller samples.  Rom\'{a}n-Z\'{u}\~{n}iga et al.~(2007)
have recently shown a similar relation among seven embedded clusters
in the Rosette nebula, which they attribute to the effects of
RGE.  In this case, the clusters are expected to have ages less than
$\sim5$~Myr, and hence should not have had a significant amount of
mass loss due to stellar evolution.  Additionally, in a sample of
young extra-galactic clusters, Ma\'iz-Apell\'aniz~(2001) found a
relation between the size of a cluster and its age, which he
attributed mainly to stellar evolutionary mass-loss.  Comparison of
detailed $N$-body models with observations of the Orion Nebular cluster
also led Scally, Clarke, \& McCaughrean~(2005) to suggest that,
despite its young age ($\sim1.5$~Myr), this cluster was substantially
more dense in the past.  Figer~(2008) has estimated the density of young massive clusters in the Galaxy, and using his data (excluding the Galactic Centre cluster) it is clear that there is a strong trend of decreasing density with increasing age, consistent with the findings of the current study.  Brandner~(2008) also has noted that young clusters in the Galaxy have larger sizes at higher ages. Finally, we note that  Scheepmaker et al.~(2007) found larger sizes for red (presumably older) clusters in the disk of M51 than blue clusters, however precise age dating of the clusters was not available.

\section{Possible causes}
\label{sec:causes}
 
As mentioned in \S~1, there are a number of possible causes for the
expansion of cluster cores with age.  We limit our discussion here to causes
that operate on the early evolution of clusters ($\lesssim 100$~Myr).

\subsection{Expansion by dynamical heating due to stellar mass black holes}
Merritt et al.~(2004) and Mackey et al.~(2007, 2008) have suggested that the
presence of stellar mass black holes in star clusters can lead to the expansion of
the core radius.  The stellar mass black holes form a dynamically distinct
(invisible) `core' and transfer energy into a stellar `halo' causing
the halo to expand, thus increasing the observed (i.e. stellar) core radius.
Merritt et al. (2004) explain the spread in the observed core radii
with age in  the LMC/SMC data of Mackey \& Gilmore~(2003) by
effectively changing the initial size scale of the cluster (through
changing the scaling to $N$-body units).  Mackey at al.~(2007, 2008) can
explain the same spread by  introducing different degrees of initial
mass segregation into their clusters and by changing the fraction of
black holes that are retained by the cluster (ie. not lost due to
large natal kick velocities).

\subsection{Stellar evolution}
When a star cluster loses mass, it will expand in an attempt to
regain virial equilibrium. The mass loss due to stellar evolution will
therefore result in an expansion of the core during the first
$\lesssim 100\,$Myr when a large fraction ($\sim20\%$) of the initial
mass is lost. However, detailed $N$-body simulations including this
effect find that the maximum growth factor of the core radius is only
about a factor of two (e.g. Portegies Zwart et al.~1999).

However, Mackey et al.~(2007, 2008) show that the effect of stellar
evolution on the expansion of a cluster is far more significant if
primordial mass segregation is included (Mackey et al. allow their 
cluster to relax for 450~Myr before turning-on stellar evolution).
Their mass segregated clusters are initially compact, with
$\rcore\simeq0.25\,$pc at $t\simeq2-3\,$Myr, which lies nicely on our
empirical fit in Fig.~\ref{fig:core-evo}. Due to the high fractional
mass loss by stellar evolution in the core, the value of $\rcore$, in the simulations of Mackey et al.~(2007,2008),
increases with log(age) roughly as $\rcore=2\log(\rm age)-1$, which
resembles our empirical curve $\rcore=1.4\log(\rm age)-0.25$. If we
attribute the core expansion as observed in Fig.~\ref{fig:core-evo}
entirely to stellar evolution, it implies that all of the clusters we
observe started with a strong degree of mass segregation. Gaburov \&
Gieles~(2008) note that $\rcore$ of mass segregated star clusters  {\it
appear} to increase with age by roughly a factor of two, due to the
massive stars that populate that core at young ages (this effect is
also included by Mackey et al.~2007, 2008).

\subsection{Residual gas expulsion (RGE) }

Clusters initially contain a significant gas fraction which is
expelled by feedback from the most massive stars after a few~Myr.  The
rapid change in the potential of the cluster causes the cluster to
expand, and possibly be destroyed (see Goodwin~2008 and references therein).  
Simulations show that $\rcore$ will expand by a factor of 5 --
10 over $\sim 10$~Myr as the cluster attempts to regain virial
equilibrium (see esp. Kroupa et al.~2001; Goodwin \& Bastian~2006;
Baumgardt \& Kroupa~2007).  The $\rcore$ evolution
of unbound clusters is very similar to the empirical fit  of
Fig.~\ref{fig:core-evo}, i.e. $\rcore$ increases linearly with
log(age).  Baumgardt \& Kroupa (2007) find clusters that remain bound
after an expansion of a factor of $\sim5$, making RGE a plausible 
explanation for the observed increase in $\rcore$.   However, Baumgardt \& Kroupa~(2007) defined their core radii in terms of the Lagrangian radii, which contains a fixed fraction of the total mass, as compared to our method which defines the core radius in terms of a profile fit.  Therefore a direct comparison between the works should be taken with caution.

For the clusters that remain bound after RGE, $\rcore$ reaches a maximum and then decline after
RGE.  The reasons for this are twofold.  Firstly,
clusters tend to `overshoot' in their attempt to re-virialise and
oscillate around virial equilibrium.  Thus the $\rcore$ are
sometimes larger than for a virialised cluster.  Secondly, $\rcore$ as measured from observations will tend
to overestimate the final (`true') $\rcore$.  After RGE a
cluster will lose a (significant) fraction of its stars even if a
bound cluster remains at the end (`infant weight loss', see e.g. 
Goodwin \& Bastian~2006).  However, stars escape at a finite speed and
so will be physically associated with the cluster for several~Myr (as
appears to be observed in a number of clusters as an excess of light
at large radii, see Bastian \& Goodwin~2006).  Thus an observer may
fit a profile that over-estimates $\rcore$ for the final, luminosity profile of the equilibrium 
cluster (Goodwin \& Bastian~2006).


\subsection{A combination of effects}

Stellar evolution and an associated expansion in the core radius {\em
must} occur in young clusters.  However, how effective this is is 
clearly highly dependent on the degree of mass segregation present in
the cluster at the onset of massive star death.

Similarly, RGE {\em must} occur as after a few~Myr
clusters change from being embedded to naked.  However, the
effectiveness of gas expulsion depends significantly on the dynamical
state of the cluster at the onset of gas expulsion, a factor for which
we have very few observational or theoretical constraints (Goodwin~2008).

The presence of a significant `dark' component in clusters as required
for later dynamical expansion is difficult to determine
observationally.  It seems plausible that at least some of the massive
stellar remnants from early stellar evolution should remain in the
cluster, but the numbers and their dynamical importance are unclear.
It should be noted that this is the only mechanism so far proposed
that can explain the later ($>100~$Myr) expansion in core
radii seen in the LMC/SMC.

Thus at least two of these proposed causes must be at work in causing
the increase in $\rcore$ with time, and probably all three (and
possibly other, as yet unknown mechanisms).  In a future paper we will theoretically
investigate the causes of core expansion in detail.

\section{Discussion and Implications}
\label{sec:discussion}


\subsection{Effect on dynamical age estimates}
The results presented in Fig.~\ref{fig:core-evo} show that estimates
of the dynamical age of a cluster, which can be defined as the number
of {\it core} relaxation times ($t_{\rm rel}$) that have passed, will be wrong when using the
current $\rcore$. Because $\rcore$ was smaller in the past, the
cluster has dynamically evolved more than one would infer from the
current properties (e.g. Portegies Zwart \& Chen 2006). Using the 
empirical fit displayed in
Fig.~\ref{fig:core-evo} we can estimate a correction factor $F$, that
is, the ratio of the true dynamical age over the dynamical age
assuming that $\rcore$ has been constant. Here we define the dynamical
age as the number of core relaxation times that have passed, so that
$F\equiv N_{\rm trel, real}/N_{\rm trel, current}$.

\begin{figure}
\includegraphics[width=9cm]{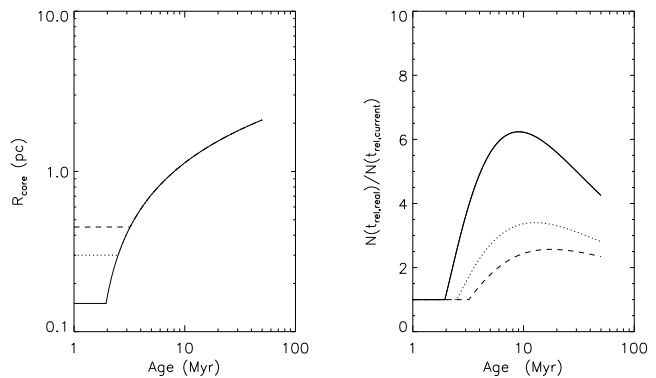}
\caption{{\bf Left:} The evolution of \rcore\ for three different
  initial radii, using the functional fit to the data in
  Fig.~\ref{fig:core-evo}.  {\bf Right:}  The ratio of N(\trelaxreal)
  (the real number of relaxation times that have actually passed) to
  N(\trelaxnow) (the number of relaxation times that have have passed
  assuming that the current relaxation time has been constant
  throughout the life of the cluster).  Depending on the initial
  radius and age of the cluster, using \trelaxnow~ significantly
  underestimates the number of relaxation times that have elapsed
  within the cluster.} 
\label{fig:trel}
\end{figure}

The core relaxation time scales as $R^{3/2}_{\rm core}$ so we can
calculate $F$ as 
\begin{equation}
F(t)=\frac{\int_0^{t} \left[\rcore(t^\prime)\right]^{3/2}{\rm d} t^\prime}{t\,R^{3/2}_{\rm core}(t)},
\label{eq:f}
\end{equation}
where we use $\rcore(t^\prime)=1.4\,\log(\rm age)-0.25$
(Fig.~\ref{fig:core-evo}). Since the empirical fit goes to $-\infty$
at $t=0$, we have to assume an initial $\rcore$ at $t=0$. In
Fig.~\ref{fig:trel} we show in the left panel the functional form of
the empirical fit, for three initial $\rcore$. In the right panel we
show the resulting $F(t)$ that follows from a numerical integration of
Eq.~\ref{eq:f}. $F$ depends strongly on the initial $\rcore$, but we
can safely say that for the observed value of very young clusters, $F$
is somewhere between $3$ and $5$ at its peak value at an age of
$\sim10\,$Myr.  For ages $\lesssim2\,$Myr $F=1$ because we have
assumed a constant $\rcore$ equal to the initial $\rcore$ there. $F$
decreases again for ages $\gtrsim10\,$Myr because the increase of
$\rcore$ has slowed down.

The results above and those shown in Fig.~\ref{fig:trel} are also valid for the {\it core} crossing time of a cluster in the limit that no stellar mass is lost during the expansion.  However, if mass loss is included the effect would be stronger on the crossing time (since $t_{\rm cross} \propto M^{-0.5}$) and weaker on the relaxation time (since $t_{\rm relax} \propto M^{0.5}$).  If the core would lose 50\% of its mass during the expansion phase, then $t_{\rm cross}$ would increase by a factor of $\sqrt 2$ and $t_{\rm relax}$ would decrease by the same amount.

This effect must be taken into account when estimating the dynamical
age of a cluster, for example to see whether the degree of mass
segregation is of primordial or dynamical origin. We discuss this more
in \S~\ref{sssec:ms}.

\subsubsection{Mass segregation}
\label{sssec:ms}
Whether a cluster is mass-segregated due to dynamical effects (energy equipartition), or if it is
primordial (set by the star/cluster formation process) in nature 
has potentially large ramifications for the
star/cluster formation process.  In order to test if a cluster's
observed mass segregation is dynamical or primordial in nature, a
comparison is often made to the observed (current) relaxation time,
\trelaxnow, to that of the cluster age.  If \trelaxnow~is greater
than the cluster age, then the mass segregation is thought to be 
primordial (e.g.~Hillenbrand  \& Hartmann~1998; Gouliermis et
al.~2004; de Grijs et al.~2002;  Chen, de Grijs, Zhao~2007). 

However, the results shown here indicate that the cluster cores expand
rapidly during the first $20$~Myr or so, and hence clusters were more
compact in the past.  Thus, \trelaxnow~may overestimate the initial
(and at earlier times) relaxation time by a large factor.  Figure~\ref{fig:trel} shows an example of this effect, although we note that these calculations are meant as an illustrative example only, as we have not included mass loss.
Indeed, Portegies Zwart \& Chen (2006) find that the (half-mass)
relaxation time can change by a factor of several due to stellar
evolution over the first $\sim 100$~Myr.

Depending on the initial radius and cluster age, estimating the number
of relaxation times that a cluster has gone through based on the
current relaxation time can result in errors of a factor 1.5 to 6.
Since this factor depends strongly on the initial cluster radius, and
since this is generally not known nor well constrained, it is highly
uncertain how many relaxation times a cluster has actually 
undergone.  Thus, claims of primordial mass segregation based on
\trelaxnow~ should be taken with caution.

\subsubsection{Stellar mergers}
The observed core expansion will significantly affect the internal
dynamics of the cluster, causing the relaxation time to increase
rapidly.   Freitag~(2007) estimates that the relaxation time could be
up to 20 times longer after the core expansion phase.   This implies
that dynamical mass-segregation, core-collapse, and/or stellar merging
only have a brief window in which to operate, namely the embedded
phase which lasts for 1-3~Myr.  The implications regarding stellar
mergers, and the subsequent formation of very massive stars, have been
considered in detail by Freitag et al.~(2006).  
They conclude that while the very dense state of the cluster
may only last for a short time, this may be compensated by the
initially very high densities.

\subsection{Older compact clusters}

While not found in our literature search (with the exception of
NGC~1569B), it is possible that some clusters remain compact ($R_{\rm
  core} < 1$~pc) during their first 10-100~Myr of evolution.  This
could happen if the effective star-formation efficiency is extremely high, if
the gas-dispersal timescale is extremely long, or if the cluster stars were born with sub-virial velocities.  However, even in these
extreme cases, some expansion is expected due to stellar evolution.
It is also clear that clusters can be formed initially with large core
radii, however these clusters would be more likely to disrupt
completely (due to RGE and stellar evolution) than their more compact
counterparts, assuming that the star formation efficiency (or, more
correctly, the initial dynamical state of the cluster) does not
relate to cluster size.

\section{Conclusions}
\label{sec:conclusions}

We have presented high S/N spectra and high-resolution imaging of six
clusters in M51.  By comparing the H$\gamma$ and H$\beta$ lines to
template spectra, we have derived their ages.  Additionally, we have
measured their structural properties using the {\it ISHAPE} code of
Larsen~(1999).  We find that the clusters are $\sim3$ to 25~Myr old
and have core radii ranging from $< 0.4$ to 1.6~pc. 

We note a strong trend between the core radius and age of the
clusters, in the sense that older clusters are larger.  Including
clusters with measured ages and structural parameters from the
literature, we find this to be a common feature in cluster evolution.
The most promising explanation of this phenomenon is that clusters
expand as they leave their embedded phase, due to the change of
gravitational potential within the cluster.  The growth in cluster 
size appears to begin at 2-3~Myr, in
good agreement with the expected/observed duration of the embedded
phase of cluster evolution and the onset of gas expulsion.
 As a cluster
expands (in particular its core) the relaxation time increases
dramatically (Freitag~2007), which limits dynamical mass segregation and significantly lowers the chances of stellar mergers (Freitag et al.~2006). 

The rather small range in mass spanned by our M51 cluster sample argues that the observed relation between age and core radius is not simply a reflection of an underlying mass-radius relation.  We caution, however, that the observed trend of increasing core radius with age could be an observational artifact if all clusters begin their lives severely mass segregated.   This would cause an underestimate of the core radius for younger clusters whose light is dominated by a few very massive stars.

These results show that the early phases of cluster evolution are
highly dynamic with many of a cluster's fundamental parameters changing
by large factors in a short time.  This leads us to caution (as did
Goodwin \& Bastian~2006) that the determination of the parameters of
young clusters must only be taken as instantaneous values, they are
not the same as a few~Myr previously, nor as they will be a few~Myr hence.

\section*{Acknowledgments}

We  thank Marc Freitag for useful
discussions  and the referee, Soeren Larsen, for his comments/suggestions which helped improve the manuscript.  NB gratefully acknowledges the hospitality of the Harvard-Smithsonian Center for Astrophysics, where a significant part of this work took place. Based on observations obtained at the Gemini Observatory, which is operated by the
Association of Universities for Research in Astronomy, Inc., under a cooperative agreement
with the NSF on behalf of the Gemini partnership: the National Science Foundation (United
States), the Science and Technology Facilities Council (United Kingdom), the
National Research Council (Canada), CONICYT (Chile), the Australian Research Council
(Australia), MinistŽrio da Cincia e Tecnologia (Brazil) and SECYT (Argentina). This paper is based on observations with the NASA/ESA {\it Hubble
Space Telescope}\/ which is operated by the Association of Universities
for Research in Astronomy, Inc. under NASA contract NAS5-26555.

\bsp
\label{lastpage}
\end{document}